\documentclass[aps,tightenlines,nofootinbib,prd]{revtex4}
\pdfoutput=1
\usepackage{amsmath,amscd,amssymb}
\usepackage{color,epsfig}
\usepackage{xfrac}
\usepackage{latexsym}
\usepackage{mathrsfs,bigints}
\usepackage{graphicx}
\usepackage{bbm}      
\usepackage{diagbox}

\newcommand{\imu}{{\rm i}}

\newcommand{\dslash}{\partial \hskip -0.5em /}

\usepackage[dvipsnames]{xcolor}

\begin{document}

\title{Excited fermions on kinks and the Dirac sea}

\author{Danial Saadatmand$^{a), b)}$, Herbert Weigel$^{a)}$}
\affiliation{$^{a)}$Institute for Theoretical Physics, Physics Department, 
Stellenbosch University, Matieland 7602, South Africa\\
$^{b)}$National Institute for Theoretical and Computational Sciences (NITheCS) South Africa}

\begin{abstract}
We study quantum effects of recently discovered kink solitons which are constructed self-consistently
by coupling to a single, excited fermion bound state. Our studies are based on the observation that 
in a semi-classical expansion the energies of this single level and of the Dirac sea should be 
treated equally. For these kink solutions we compute the energy of the Dirac sea as the fermion 
vacuum polarization energy. We find it to be substantial and to typically outweigh the energy gain 
from binding the single level.
\end{abstract}

\maketitle

\section{Introduction and Motivation}
\label{sec:intro}
By Derrick's theorem \cite{Derrick:1964ww} scalar field theories in one time and one space 
dimensions ($D=1+1$) wherein spontaneous symmetry breaking produces discrete, degenerate vacua 
are almost certain to contain (static) soliton solutions. These solutions have localized
energy densities \cite{Ra82,Manton:2004tk,Vachaspati:2006zz}. Low dimensional soliton models
are frequently considered as toy models for more complex theories in higher dimensions that have 
applications in many branches of physics: in cosmology~\cite{Vilenkin:2000jqa}, condensed matter 
physics~\cite{Schollwock:2004aa,Nagasoa:2013}, as well as hadron \cite{Weigel:2008zz} and nuclear 
physics~\cite{Feist:2012ps}.

The field equation for a soliton is 
equivalent to minimizing its classical energy, $E_{\rm cl}$. The leading, one-loop quantum correction 
to $E_{\rm cl}$ is the renormalized sum of the shifts of the zero point energies of the quantum 
fluctuations. These shifts reflect the polarization of the vacuum induced by the soliton and thus this 
quantum correction is frequently called the vacuum polarization energy $E_{\rm VPE}$ (VPE). By now 
techniques have been developed that make it relatively straightforward to compute the VPE in 
$D=1+1$ \cite{Weigel:2016zbs,Graham:2022rqk}; this is particularly the case when the potential 
for the quantum fluctuations is reflection symmetric. In general, $E_{\rm cl}$ and $E_{\rm VPE}$ exhibit 
different dependences on the model parameters. This feature is, for example, fundamental for the soliton 
picture of baryons in effective theories for quantum chromodynamics~\cite{Witten:1979kh}. If the model 
parameters are chosen such that $E_{\rm cl}\approx E_{\rm VPE}$ one would expect that higher loop 
corrections are not negligible.\footnote{See {\it e.g.} Ref. \cite{Evslin:2021vgk} for estimates beyond 
one loop.} In such cases the VPE is not a reliable approximation for the quantum correction and its 
computation is merely of academic interest. Of course, this observation does not at all lessen the
value of the ground breaking studies on the VPE about half a century ago \cite{Dashen:1974ci}.

Yet, there are scenarios in which the VPE is indeed of significant importance and must be included. 
First, $E_{\rm cl}$ may be degenerate with respect to a certain (variational) parameter for the soliton but 
$E_{\rm VPE}$ is not.\footnote{Translational invariance usually holds for both, though there are exceptions 
\cite{Weigel:2016zbs}.} In that case the VPE is decisive for the favorable soliton configuration albeit
$E_{\rm cl}\gg E_{\rm VPE}$. Ref.~\cite{Graham:2022rqk} explores cases in which the inclusion of the 
VPE even destabilizes classically stable solitons (Higher loop corrections may reverse this picture.).
Second, the field equations are usually derived by minimizing an energy functional. This functional
may contain components that in some expansion scheme are of the same order as the VPE. This feature can 
be delicate because the particular parameter dependences may be hidden when the model is constructed in 
terms of dimensionless variables and coordinates. It is this scenario that we focus on in this study. Third, 
with different topological sectors and associated topological charges $Q$, the VPE may be decisive for 
binding energies as in $E(2Q)-2E(Q)$ \cite{Graham:2022adn}.

The second scenario described above applies to models that recently attracted renewed interest. In these models 
a scalar boson with a non-linear self-interaction that allows spontaneous symmetry breaking has a Yukawa coupling 
with a fermion. Typically the scalar field assumes a kink type structure connecting different vacua at positive 
and negative spatial infinity (and is thus topologically stable). Novel local minima of the energy functional were 
obtained \cite{Klimashonok:2019iya,Perapechka:2019vqv,Gani:2022ity} when the scalar couples to a single fermion 
that dwells in an exited bound state. We call these configurations local minima because the fermion could decay 
into a lower energy state while the kink radiates small amplitude fluctuations without changing its topological 
structure. The resulting configuration would have a smaller total energy and the classical kink mass would be 
its lower bound. We want to revisit these local minima for two reasons. First, in their construction the Dirac 
sea contribution was omitted. As we will explain later, in any suitable expansion scheme this contribution to the 
energy is of the same order as the one from the single level. Second, solutions with the single fermion occupying 
a negative energy level were also constructed. The Dirac sea must be included to give a physical interpretation 
because the negative energy level is a hole in the sea. But then the solution to the field equation becomes 
the charge conjugation of a configuration where the boson couples to a positive energy fermion level. 

The paper is organized as follows. In section \ref{sec:sea} we will give a formal and brief discussion on 
the role of the Dirac sea when a single fermion level is occupied. We will then review the construction 
procedure of Ref. \cite{Klimashonok:2019iya} modified such that the hole character of the negative energy
levels is respected. In section \ref{sec:vpe} we will list and explain the formulas relevant to compute the 
vacuum polarization energy in the no-tadpole renormalization scheme. We present the results of our numerical 
simulations in section \ref{sec:numres} and conclude with a summary in section \ref{sec:conclude}.

\section{Dirac sea}
\label{sec:sea}

In order to make the arguments from the introduction explicit, we start with a short discussion of the 
role of the Dirac sea when certain fermion levels with energy eigenvalues $E_\nu$ are occupied. For a charge 
conjugation invariant background we can formally, {\it i.e.} before regularization and renormalization, write 
the total energy as
\begin{equation}
E_{\rm tot}=E_{\rm cl}+\sum_\nu \eta_\nu |E_\nu| 
-\sum_\nu \left[\theta(-E_\nu)|E_\nu|-\theta(-E^{(0)}_\nu)|E^{(0)}_\nu|\right]\,.
\label{eq:etot1}\end{equation}
Here $E_{\rm cl}$ is the classical energy of the scalar background and $\eta_\nu=0,1$ are occupation numbers 
that describe the occupation of particular levels. The second sum adds the Dirac sea 
contribution\footnote{If the background was not charge conjugation variant we would write $\frac{1}{2}\sum_\nu |E_\nu|$ 
for the Dirac sea contribution. Even though the trace of the Dirac Hamiltonian vanishes formally, 
the two prescriptions may yield different results because particular regularization prescriptions
may cause this trace to be no longer zero.} where we have subtracted the trivial background equivalent 
(indicated by the superscript). The difference in that second sum measures the change in the 
spectrum and is called the vacuum polarization energy~(VPE).

Typically only a single level is selected, call it $n$ so that $\eta_n=1$ and $\eta_{\nu\ne n}=0$. We classify a 
configuration as {\it classically stable} if $E_{\rm cl}+|E_n|\le m$, where $m$ is the mass of the Dirac field. 
The {\it kink} is the soliton for a model without fermions. Its classical energy is $E_{\rm kink}$. We call a 
configuration whose total energy is $E_{\rm tot}\le m+E_{\rm kink}$ {\it topologically stable} because the sum on 
the right hand side is the minimal energy of the system of an isolated free fermion and a boson field with a non-trivial 
topological structure. This is a sensible comparison because the change required for the boson field to get from $E_{\rm cl}$
to $E_{\rm kink}$ does not alter that structure. 

When $E_n>0$ we simply occupy that level. It is more interesting to consider
$E_n<0$. Then we can write
$$
|E_n|-\sum_\nu\left[\theta(-E_\nu)|E_\nu|-\theta(-E^{(0)}_\nu)|E^{(0)}_\nu|\right]
=-\sum_{\nu\ne n} \theta(-E_\nu)|E_\nu|+\sum_\nu\theta(-E^{(0)}_\nu)|E^{(0)}_\nu|\,,
$$
which means that we have created a hole in the Dirac sea corresponding to an anti-particle state. Obviously
the inclusion of the Dirac sea is essential for a consistent particle or anti-particle interpretation of
the solutions to the Dirac equation. The need for combining the level and sea contributions was 
actually noted quite early in the context of non-topological soliton models for baryons 
\cite{Friedberg:1976eg,Kahana:1984dx}, though those studies only focused on the case when the lowest 
non-negative energy bound state was occupied. Also renormalization is an issue for those higher 
dimensional models.

In our analysis we first follow the procedure of Ref.~\cite{Klimashonok:2019iya} and minimize $E_{\rm cl}+|E_n|$ 
to construct the static kink profile~$\Phi_n$. We will subsequently compute the Dirac sea contribution for this 
kink profile. In principle the Dirac sea component must also be included when constructing the profile from the 
minimum condition for the total energy. However, that is quite complicated as highly non-local field equations 
(would) emerge. So far this has only been performed in variational approximations or in non-renormalizable theories, 
{\it cf.} Refs.~\cite{Graham:2009zz,Alkofer:1994ph} for reviews. 

\section{The Kink Model}
\label{sec:model}
In $D=1+1$ the scalar field $\Phi$ is dimensionless while the fermion spinors $\Psi$ have
canonical energy dimension $\frac{1}{2}$. To make the Yukawa coupling constant $g$
dimensionless we write the Lagrangian as\footnote{For quantized fermion fields we should actually 
write the fermion part of the Lagrangian as
$$
{\cal L}_F=\frac{\imu}{2}\left[\bar{\Psi},\dslash\Psi\right]
-\frac{g}{2}\sqrt{\frac{\lambda}{2}}\,\Phi\left[\bar{\Psi},\Psi\right]
$$
to properly account for charge conjugation. This gives rise to the factor ${\rm sign}(E_n)$ in 
Eq.~(\ref{eq:eqm1}). The commutators are for the fermion creation and annihilation operators.}
\begin{equation}
\mathcal{L}=\frac{1}{2}\partial_\mu\Phi\partial^\mu\Phi
-\frac{\lambda}{4}\left(\Phi^2-\frac{M^2}{2\lambda}\right)^2
+\imu \overline{\Psi}\dslash\Psi-g\sqrt{\frac{\lambda}{2}}\,\overline{\Psi}\Phi\Psi\,.
\label{eq:lag}\end{equation}
Observe that $\lambda$ has dimension energy squared and that $m=\frac{gM}{2}$ is the fermion mass from spontaneous symmetry 
breaking that generates the vacuum expectation value $\langle \Phi \rangle=\frac{\pm M}{\sqrt{2\lambda}}$. The 
field equations for the scenario in which the scalar field only couples to the level $n$ are
(with the convention $\gamma^0=\sigma_1$ and $\gamma^1=\imu\sigma_3$)
\begin{align}
\partial_x^2\Phi_n&=\lambda\Phi\left(\Phi_n^2-\frac{M^2}{2\lambda}\right)
+g\sqrt{\frac{\lambda}{2}}\,{\rm sign}(E_n)\,\Psi^\dagger_n\sigma_1\Psi_n\cr
E_n\Psi_n&=-\imu\sigma_2\partial_x \Psi_n +g\sqrt{\frac{\lambda}{2}}\,\Phi_n\sigma_1\Psi_n\,.
\label{eq:eqm1} \end{align}
These equations are supplemented by the normalization condition $\bigintsss dx \, \Psi_n^\dagger\Psi_n=1$.
Accounting for the fermion source term in the scalar field equation has been phrased 
{\it back-reaction} in Ref. \cite{Klimashonok:2019iya}.

In order to find the most generic, {\it i.e.} parameter independent, formulation and 
also for numerical practicality it is appropriate to introduce dimensionless quantities:
\begin{equation}
\Phi_n(x)=\frac{M}{\sqrt{2\lambda}}\,\phi(\xi)
\quad {\rm and}\quad 
\Psi_n(x)=\sqrt\frac{M}{2}\,\psi(\xi)
\quad {\rm where}\quad 
\xi=\frac{M}{2}x\,.
\label{eq:scale1}\end{equation}
For simplicity we omit the subscript on the new fields. After this transformation the normalization 
condition is $\bigintsss d\xi\, \psi^\dagger(\xi)\psi(\xi)=1$. The field equations become
\begin{equation}
\phi^{\prime\prime}(\xi)=2\phi(\xi)\left(\phi^2(\xi)-1\right)
+g\frac{2\lambda}{M^2}{\rm sign}(\epsilon)\psi^\dagger(\xi)\sigma_1\psi(\xi)
\quad{\rm and}\quad
\epsilon \psi(\xi)=-\imu\sigma_2\psi^\prime(\xi)+g\phi(\xi)\sigma_1\psi(\xi)\,,
\label{eq:scale2}\end{equation}
where $\epsilon=\frac{2E_n}{M}$ is also dimensionless and primes denote derivatives with respect
to $\xi$. The dimensionless fermion mass in this parameterization is $m\,\sim\,g$.

The classical mass is the integral
\begin{equation}
E_{\rm cl}=\int dx\,\left[\frac{1}{2}\left(\partial_x \Phi(x)\right)^2
+\frac{\lambda}{4}\left(\Phi^2(x)-\frac{M^2}{2\lambda}\right)^2\right]
=\frac{M^3}{8\lambda}\int d\xi\,\left[\phi^{\prime2}(\xi)+\left(\phi^2(\xi)-1\right)^2\right]
=\frac{M^3}{\lambda}\,\frac{\epsilon_{\rm cl}}{4}\,,
\label{eq:ecl1}\end{equation}
where we understand the last equation as the definition of $\epsilon_{\rm cl}$. Then we combine
\begin{equation}
E_{\rm cl}+|E_n|=
\frac{M^3}{4\lambda}\left[\epsilon_{\rm cl}+\frac{2\lambda}{M^2}\,|\epsilon|\right]\,.
\label{eq:ecl2}\end{equation}
As in Ref. \cite{Friedberg:1976eg} we call this the {\it quasi-classical energy}. We make the important 
observation that the contribution from the fermion levels scales with the relative factor $\frac{2\lambda}{M^2}$. 
This, of course, is then also the case for the energy from the Dirac sea because that is the sum of the single fermion 
energies obtained from the very same wave-equation. This different parameter dependence is not surprising since 
in a semi-classical expansion, the VPE has an extra factor of~$\hbar$. Apart from the above particle-hole 
interpretation we have thus another strong motivation to include the Dirac sea. In Ref.~\cite{Klimashonok:2019iya} 
only the particular parameter choice $M=\sqrt{2\lambda}$ was considered. To some extent this hides the crucial 
detail that the classic scalar and fermion quantum energies have different scaling behaviors.

Introducing upper and lower spinor components $\psi=\begin{pmatrix}u\\[-1mm]v\end{pmatrix}$,
the profiles are subject to the differential equations
\begin{equation}
\phi^{\prime\prime}=2\phi\left(\phi^2-1\right)+\frac{4g\lambda}{M^2}\,{\rm sign}(\epsilon)uv\,,\qquad
u^\prime=\epsilon v -g\phi u \qquad {\rm and}\qquad
v^\prime=-\epsilon u +g\phi v\,.
\label{eq:scale3}\end{equation}
For $g=0$ the soliton is the well-known {\it kink}, $\phi(x)=\tanh(x)$.
For completeness we recall that one then finds $\epsilon_{\rm cl}=\frac{4}{3}$, or 
$E_{\rm cl}=E_{\rm kink}=\frac{M^3}{3\lambda}$. The kink is odd under spatial reflection.
This reflection property is maintained when $g$ is smoothly increased and we can write
\begin{equation}
\epsilon_{\rm cl}=\int_0^\infty d\xi\,\left[\phi^{\prime2}(\xi)+\left(\phi^2(\xi)-1\right)^2\right]\,.
\label{eq:scale4}\end{equation}
From $\phi(-\xi)=-\phi(\xi)$ it is furthermore obvious that there are two parity cases for the fermion spinors. 
The first has even $u$ and odd $v$; the second has even $v$ and odd $u$. These cases are called $A$-and $B$-type 
solutions in Ref. \cite{Klimashonok:2019iya}. We keep that notation. Additional integer labels on $A$ and $B$ count 
the number of zero-crossings of $u$ on the half-line $x\ge0$, including the one at $x=0$ for the $B$-type.

Eqs.~(\ref{eq:scale3}) are invariant when changing the signs of $u$ and $\epsilon$ but keeping the sign of $v$. 
Hence the spectrum is symmetric for a given profile $\phi$. Furthermore occupying a particle or an anti-particle 
level with equal $|\epsilon|$ (which is equivalent to digging a hole into the Dirac sea) yield equivalent solutions.

The fermion equations have two $\epsilon=0$ solutions 
\begin{equation}
u(\xi)=N{\rm exp}\left[-g\int^\xi d\xi^\prime \phi(\xi^\prime)\right]\,,\quad v(\xi)=0
\qquad {\rm and}\qquad
u(\xi)=0\,,\quad v(\xi)=N{\rm exp}\left[g\int^\xi d\xi^\prime \phi(\xi^\prime)\right]\,.
\label{eq:zeromode}\end{equation}
Without loss of generality we take ${\rm sign}(\phi)={\rm sign}(\xi)$. Then only the first solution is 
normalizable so that there is exactly one normalizable zero mode in the $A_0$ channel for any prescribed~$\phi$. 
In either case $uv=0$ so that there is no fermionic source term in the wave-equation for $\phi$ which is 
then solved by the kink profile.

\section{Fermion VPE}
\label{sec:vpe}

The construction of fermion continuum scattering solutions with a kink type background is not
straightforward because the mass parameters at positive and negative spatial infinity have 
opposite signs; though this is only a problem for practical calculations rather than a conceptual
one. Fortunately for a static system which is invariant under charge conjugation we can formally 
write the effective action, from which the VPE is extracted, as
\begin{equation}
\int \frac{d\omega}{2\pi}\,{\rm Tr}{\rm Log}\left[\omega-H\right]
=\frac{1}{2}\int \frac{d\omega}{2\pi}\,{\rm Tr}{\rm Log}\left[\omega^2-H^2\right]\,,
\label{eq:Aeff}\end{equation}
where the trace goes over the eigenstates of the Dirac Hamiltonian $H$. This shows that the VPE of the fermion 
system can be obtained from the average VPE of two scalar systems \cite{Graham:1999pp}. More explicitly we 
have from Eq.~(\ref{eq:scale3})
\begin{equation}
H=\begin{pmatrix}0 & -\partial_\xi +g\phi \cr \partial_\xi +g\phi & 0 \end{pmatrix}
\qquad \mbox{so that}\qquad
H^2=\begin{pmatrix} -\partial^2_\xi -g\phi^\prime+g^2\phi^2 & 0 \cr 
0 & -\partial^2_\xi +g\phi^\prime+g^2\phi^2 \end{pmatrix}\,.
\label{eq:vpe1}\end{equation}
The potentials of the two scalar systems are straightforwardly identified as 
\begin{equation}
V_S=g^2\left(\phi^2-1\right)-g\phi^\prime
\qquad \mbox{and}\qquad
\widetilde{V_S}=g^2\left(\phi^2-1\right)+g\phi^\prime\,.
\label{eq:vpe2}\end{equation}
With $\phi$ being odd, these potentials are even in the coordinate $\xi$ and the VPE in the no-tadpole 
renormalization scheme can be computed by standard techniques \cite{Graham:2009zz,Graham:2022rqk}. In this 
scheme, the Lagrangian counterterm proportional to $\left(\Phi^2-\langle\Phi\rangle^2\right)$ is adjusted such 
that there are no quantum corrections to the vacuum expectation value of the scalar field. The related, 
ultraviolet divergent tadpole Feynman diagram is a loop with a single insertion of the potential and 
thus does not depend on the Fourier momentum of $V_S$ (or $\widetilde{V_S}$). In turn the diagram
is completely canceled by this counterterm. 

The basic ingredients for the non-perturbative part of the VPE are the Jost functions for imaginary momenta 
in the positive and negative parity channels. The Jost functions are extracted from the Jost solutions $f(k,\xi)$ 
which are solutions to the Schroedinger equations with the potentials $V_S$ and $\widetilde{V_S}$. Asymptotically 
($\xi\to\infty$) the Jost solutions are incoming plane waves. In general there are contributions to the VPE from 
the bound and continuum scattering states. The latter contribution is the momentum integral over the single particle 
energies weighted by the density of states. In a given parity channel the difference of the densities of states 
with and without the background potential is the derivative of the phase shift with respect to the 
momentum~\cite{Faulkner:1977zz}. Next we use that the phase shift is the logarithm of the 
Jost function which has simple zeros at the imaginary momenta of the bound state energies. 
Hence when writing that continuum integral as a contour integral in the complex momentum plane the singularities 
and residues resulting from the logarithmic derivative cancel the explicit bound state contributions to the 
VPE. All what is left stems from the discontinuity of the dispersion relation $\omega=\sqrt{k^2+g^2}$
along the imaginary axis.\footnote{The imaginary axis formalism also avoids ambiguities from the 
multivalued logarithm of complex numbers that might have occurred in earlier studies on the fermion
VPE from (pseudo)scalar backgrounds \cite{Graham:1999pp,Farhi:2000ws,Gousheh:2012ec}.} With $k=\imu t$ we 
are then left with an integral over $t\in\left[g,\infty\right]$. 

To facilitate the calculation we
factorize the plane wave component of the Jost solution after continuing to imaginary momenta: 
$f(\imu t,\xi)={\rm e}^{-t\xi}G(t,\xi)$ and establish the second order differential equation
\begin{equation}
G^{\prime\prime}(t,\xi)=2tG^\prime(t,\xi)+\sigma(\xi)G(t,\xi)\,.
\label{eq:vpe3}\end{equation}
Here $\sigma$ is either $V_S$ or $\widetilde{V_S}$.
With the boundary condition $\lim_{\xi\to\infty}G(t,\xi)=1$ we extract the Jost functions from
$G(t,0)$ and $G^\prime(t,0)=\frac{\partial G(t,\xi)}{\partial \xi}\Big|_{\xi=0}$ to accommodate the
reflections properties (of the scattering wave-function) for the positive and negative parity channels. 
Putting things together yields the vacuum polarization energy \cite{Graham:2009zz,Graham:2022rqk}
\begin{equation}
\epsilon_{\rm VPE}[\sigma]=\int_{0}^\infty \frac{d\tau}{2\pi}\,\left\{
{\rm ln}\left[G(t,0)\left(G(t,0)
-\frac{1}{t}G^\prime(t,0)\right)
\right]-\frac{\langle \sigma\rangle}{t}\right\}_{t=\sqrt{\tau^2+g^2}}\,.
\label{eq:vpe4} \end{equation}
Here we have introduced the integration variable $\tau=\sqrt{t^2-g^2}$ to mitigate an integrable singularity 
at $t=g$. The two factors under the logarithm are the Jost functions in the odd and even parity channels, 
respectively while the subtraction with $\langle \sigma\rangle=\bigintsss_0^\infty d\xi\,\sigma(\xi)$ takes out 
the contribution that is associated with the tadpole diagram. Since that diagram and the counterterm cancel exactly,
the subtraction in Eq.~(\ref{eq:vpe4}) implements the no-tadpole renormalization condition and renders the $\tau$ 
integral ultra-violet finite. The total renormalized VPE is simply the sum (note the overall sign for a fermion)
\begin{equation}
E_{\rm VPE}=-\frac{M}{4}\left(\epsilon_{\rm VPE}\left[V_S\right]
+\epsilon_{\rm VPE}\left[\widetilde{V}_S\right]\right)\,.
\label{eq:vpe5} \end{equation}
For this average the subtraction under the integral in Eq.~(\ref{eq:vpe4}) is proportional to
\begin{equation}
\frac{1}{2}\left\langle V_S + \widetilde{V}_S\right\rangle=g^2\int_0^\infty d\xi\, \left(\phi^2-1\right)\,.
\label{eq:tadpole}\end{equation}
It indeed equals the contribution from the counterterm that cancels all quantum corrections to the 
scalar vacuum expectation value.

\section{Numerical Results} 
\label{sec:numres}
The numerical treatment Eqs.~(\ref{eq:scale3}) is simplified by the parity properties that we have discussed 
in Sect.~\ref{sec:model}. For a given parity channel it suffices to solve the equations on the half-line
$\xi\ge0$. In addition to $\phi(0)=0$ we have the following initial conditions for the spinors
\begin{equation}
\mbox{A-type:}\quad u(0)=1\,,\quad v(0)=0
\qquad {\rm and}\qquad
\mbox{B-type:}\quad u(0)=0\,,\quad v(0)=1\,.
\label{eq:init}\end{equation}
At a large $\xi_{\rm max}$ in both channels the boundary conditions 
\begin{equation}
u(\xi_{\rm max})=1\qquad {\rm and}\qquad
v(\xi_{\rm max})=\frac{g-t}{\epsilon}
\qquad {\rm with}\qquad t=\sqrt{g^2-\epsilon^2}
\label{eq:bc}\end{equation}
ensure that the fermion profiles are proportional to ${\rm e}^{-t\xi}$ 
outside the realm of $\phi(\xi)$ as required for a bound state with $|\epsilon|<g$.

To self-consistently construct the profile that minimizes the quasi-classical energy 
$\epsilon_{\rm cl}+\frac{2\lambda}{M^2}|\epsilon|$ we first compute the bound state energies 
and wave-functions as a function of $g$ for the kink profile $\phi(x)=\tanh(x)$. In this process 
the fermion differential equations in Eq.~(\ref{eq:scale3}) are solved with the initial and 
boundary conditions from Eqs.~(\ref{eq:init}) and~(\ref{eq:bc}). Solutions that are continuous over 
the whole $\xi$ half-line only exist for particular values of $\epsilon$. These are the energy eigenvalues 
which we identify with a nested interval algorithm. The resulting fermion profile functions are normalized to 
$\bigintsss_0^\infty d\xi\, \left[u^2+v^2\right]=\frac{1}{2}$. At small $g$ only the $A_0$ zero mode is bound. 
We then increase $g$ until the desired mode gets bound with a significant binding energy. This desired mode could, 
for example, be $B_1$ which is the one with the lowest positive energy eigenvalue in the negative parity channel. We 
substitute the corresponding fermion profiles into the differential equation for~$\phi$. For this equation we employ 
a shooting method for the slopes of $\phi(\xi)$ at $\xi=0$ and $\xi=\xi_{\rm max}\gg1/g$ such that both $\phi$ and 
$\phi^\prime$ are continuous. We then iterate this procedure until convergence under the iteration is observed. 
This convergence is measured by $\epsilon_{\rm cl}+\frac{2\lambda}{M^2}|\epsilon|$ being stationary.
As we further increase $g$, guessing a good initial profile $\phi$ becomes problematic as we observe that the 
success of the iteration is quite sensitive to this initial guess. In most cases 
a linear (over-)relaxation procedure has proven to be successful: Assume $\phi^{(1)}$ and $\phi^{(2)}$ are 
self-consistent solutions for Yukawa couplings $g_1\,{\scriptstyle \lesssim}\,g_2$, respectively. Then
\begin{equation}
\phi^{(3)}(\xi)=\phi^{(1)}(\xi)+\zeta\frac{\phi^{(2)}(\xi)-\phi^{(1)}(\xi)}{g_2-g_1}\left(g_3-g_1\right)\,,
\label{eq:relax}\end{equation}
with the fudge factor $\zeta\,{\scriptstyle \gtrsim}\,1$, turns out to be a good initial guess 
for the profile associated with $g_3\,{\scriptstyle \gtrsim}\,g_2$.

In all numerical simulations we set the scale by choosing $M=2$ and eventually vary $\lambda$. 
Stated otherwise, to get energies in terms of physical units, the results below have to be multiplied by $M/2$ with $M$ 
measured in electron-volts, for example. This choice for $M$ is convenient since then $E=\frac{2}{M}\epsilon=\epsilon$.

\begin{figure}
\centerline{\epsfig{file=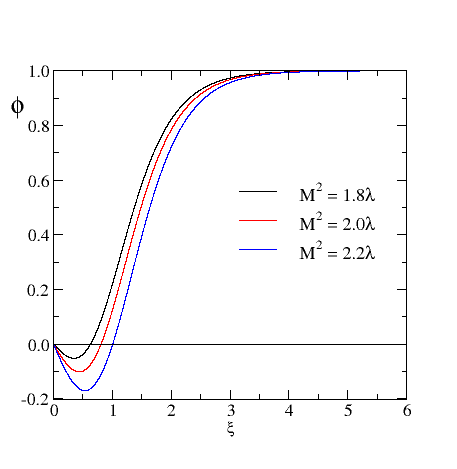,width=7cm,height=5cm}\hspace{1cm}
\epsfig{file=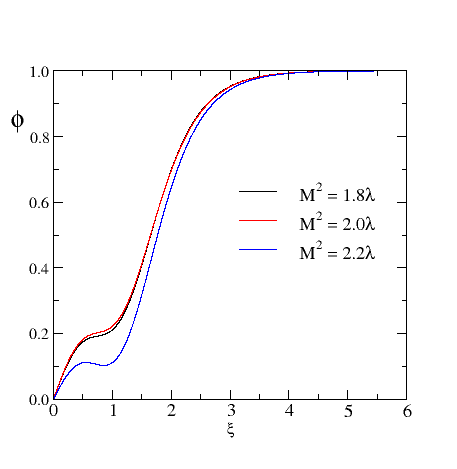,width=7cm,height=5cm}}
\caption{\label{fig:kink1}Kink profiles in the $B_1$ (left panel) and $A_1$ (right panel) channels 
for $g=4.0$ and deviations from $M^2=2\lambda$. The profile on the negative half-line is obtained from
the reflection $\phi(\xi)=-\phi(-\xi)$.}
\end{figure}

In Ref.~\cite{Klimashonok:2019iya} solutions to Eqs.~(\ref{eq:scale3}) were obtained for many choices of the 
occupied level. To make our arguments clear it suffices to focus on the lowing lying excitations in the two parity 
channels $B_1$ and $A_1$. For these channels we have reproduced the kink solitons of Ref.~\cite{Klimashonok:2019iya} when
$M^2=2\lambda$. In addition we have constructed solutions deviating from this particular relation between the 
model parameters. A few profiles are displayed in figure~\ref{fig:kink1}. Their construction turned out to be somewhat 
cumbersome for $M^2<2\lambda$ since this effectively corresponds to increasing $g$ in the differential equation for $\phi$. 
Accordingly the larger the ratio $\frac{\lambda}{M^2}$, the larger the deviation of the kink profile from 
$\tanh(\xi)$. 

For the discussion of the energetic stability we recall from the previous sections the three energies
to consider, (i)~classical: $E_{\rm cl}$, (ii) quasi-classical: $E_{\rm cl}+E$ and (iii) semi-classical:
$E_{\rm tot}=E_{\rm cl}+E+E_{\rm VPE}$. As argued in Sect.~\ref{sec:model} (i) and (iii) are leading and 
sub-leading fermion contributions, respectively, while (ii) cannot be associated with a particular truncation 
of an expansion scheme.

\begin{table}
\centerline{
\begin{tabular}{c|cccc||c|cccc}
\multicolumn{5}{c||}{$B_1$}&
\multicolumn{5}{c}{$A_1$}\cr
\hline
$g$ & $E$ & $E_{\rm cl}$ & $E+E_{\rm cl}$ & $g+E_{\rm kink}$ & 
$g$ & $E$ & $E_{\rm cl}$ & $E+E_{\rm cl}$ & $g+E_{\rm kink}$ \cr
\hline
$2.6$ & $1.433$ & $1.670$ & $3.103$ & $3.933$ &
$2.8$ & $2.227$ & $1.536$ & $3.764$ & $4.133$ \\
$3.0$ & $1.370$ & $1.812$ & $3.183$ & $4.333$ &
$3.0$ & $2.140$ & $1.721$ & $3.861$ & $4.333$ \\
$4.0$ & $1.039$ & $2.231$ & $3.270$ & $5.333$ &
$4.0$ & $2.069$ & $2.079$ & $4.148$ & $5.333$ \\
$5.0$ & $0.715$ & $2.540$ & $3.255$ & $6.333$ &
$5.0$ & $1.749$ & $2.542$ & $4.292$ & $6.333$ \\
\end{tabular}}
\caption{\label{tab:cl1}Energies in Eq.~(\ref{eq:ecl2}) as functions of the Yukawa coupling $g$
for the $B_1$ and $A_1$ channels with $M^2=2\lambda$. For $g=0$ the classical mass is 
$E_{\rm kink}=\frac{4}{3}\approx1.333$.}
\end{table}

\begin{table}
\centerline{
\begin{tabular}{c|cccc||c|cccc}
\multicolumn{5}{c||}{$B_1$}&
\multicolumn{5}{c}{$A_1$}\cr
\hline
$g$ & $E$ & $E_{\rm cl}$ & $E+E_{\rm cl}$ & $g+E_{\rm kink}$ &
$g$ & $E$ & $E_{\rm cl}$ & $E+E_{\rm cl}$ & $g+E_{\rm kink}$ \cr
\hline
$2.6$  & $1.346$  & $1.584$  & $2.930$  & $3.8$ & 
$3.0$  & $2.090$ & $1.595$  & $3.685$   & $4.2$ \\
$3.0$  & $1.227$  & $1.765$  & $2.992$  & $4.2$ & 
$3.5$  & $2.036$ & $1.799$  & $3.835$   & $4.7$ \\
$3.5$  & $1.026$  & $2.006$  & $3.032$  & $4.7$ & 
$4.0$  & $1.875$ & $2.068$  & $3.944$   & $5.2$ \\
$4.0$  & $0.812$  & $2.223$  & $3.035$  & $5.2$ &
$4.5$  & $1.716$ & $2.286$  & $4.003$   & $5.7$ \\
\end{tabular}}
\caption{\label{tab:cl2}Same as table \ref{tab:cl1} for $M^2=1.8\lambda$ with
$E_{\rm kink}=\frac{3}{5}=1.2$.}
\end{table}

To investigate the (quasi-)classical stability ($E+E_{\rm cl}\le g$) we consider the numerical 
results for Eq.~(\ref{eq:ecl2}) in tables \ref{tab:cl1}, \ref{tab:cl2} and \ref{tab:cl3}. We stress 
that this is not absolute stability, as the fermion could still move to the ever present zero mode
and the energy would be that of the classical kink, $E_{\rm kink}=\frac{M^3}{3\lambda}$ plus the
fermion VPE for the kink (to be discussed below). We observe this stability when the Yukawa coupling
$g$ exceeds a certain value. With $M^2=2\lambda$ this is at about $g=3$ and $g=4$ for the $B_1$ and
$A_1$ channels, respectively. As we detune $\lambda$ from this relation, these critical values
decrease as $\lambda$ increases and vice versa. This opposite behavior is essentially due to 
the $\lambda$ dependence of $E_{\rm cl}$. We recognize that $E_{\rm cl}$ is smaller for the $A_1$
channel than for the $B_1$ channel when $M^2\le2\lambda$. This is remarkable because the distortion
of the $A_1$ profile is larger in the sense that it is wider than its $B_1$ counterpart.

\begin{table}
\centerline{
\begin{tabular}{c|cccc||c|cccc}
\multicolumn{5}{c||}{$B_1$}&
\multicolumn{5}{c}{$A_1$}\cr
\hline
$g$ & $E$ & $E_{\rm cl}$ & $E+E_{\rm cl}$ & $g+E_{\rm kink}$ &
$g$ & $E$ & $E_{\rm cl}$ & $E+E_{\rm cl}$ & $g+E_{\rm kink}$ \cr
\hline
$3.0$  & $1.472$ & $1.887$ & $3.358$  & $4.467$ &
$3.0$  & $2.146$ & $1.887$  & $4.039$ & $4.467$  \\
$4.0$  & $1.226$ & $2.258$ & $3.484$  & $5.467$ & 
$4.0$  & $2.050$ & $2.304$  & $4.354$ & $5.467$ \\
$5.0$  & $0.891$ & $2.609$ & $3.501$  & $6.467$ &
$5.0$  & $1.813$ & $2.721$  & $4.535$ & $6.467$ 
\end{tabular}}
\caption{\label{tab:cl3}Same as table \ref{tab:cl1} for $M^2=2.2\lambda$ with
$E_{\rm kink}=\approx1.467$. }
\end{table}

We next compute the VPE. Occupying the zero-mode~$A_0$ implies $uv=0$, according to 
Eq.~(\ref{eq:zeromode}). Hence there is no back-reaction in this case and $\phi(\xi)=\tanh(\xi)$. 
For this scenario we find $E_{\rm VPE}=0.348$, $0.607$, $0.943$ and $1.359$ for $g=2.0$, $g=3.0$, 
$g=4.0$ and $g=5.0$, respectively. The $g=2.0$ result is analytically known to be
$M\left(\frac{1}{\pi}-\frac{1}{4\sqrt{3}}\right)\approx0.3479$ \cite{Graham:1999pp}
favorably confirming our numerical simulation. The VPE results for the $A_0$ level do not depend 
on the particular value for $\lambda$ since this parameter enters via the distortion of 
the scalar profile only when $uv\ne0$ and as an overall factor in $E_{\rm cl}$.
The resulting total energies are listed in table \ref{tab:lam0}. In our approximation scheme these 
are the lowest energies for the kink topology. In this channel we have $E=0$ and $E_{\rm VPE}<g$ so 
that the configuration is topologically stable.
\begin{table}
\centerline{
\begin{tabular}{c|cccc}
&\multicolumn{4}{|c}{$E_{\rm kink}+E_{\rm VPE}$}\cr
\hline
\diagbox[height=0.6cm,width=1.5cm]{\tiny\, $M^2/\lambda$}{\small~~$g$}
& $2.0$ & $3.0$ & $4.0$ & $5.0$\cr
\hline
$1.8$ &~~ $1.548$~~ &~~ $1.807$~~ &~~ $2.143$~~ &~~ $2.559$ \cr
$2.0$ &~~ $1.681$~~ &~~ $1.940$~~ &~~ $2.276$~~ &~~ $2.692$ \cr
$2.2$ &~~ $1.815$~~ &~~ $2.074$~~ &~~ $2.410$~~ &~~ $2.826$
\end{tabular}}
\caption{\label{tab:lam0}Total energy for the $A_0$ channel for various
model parameters.}
\end{table}

\begin{table}
\centerline{
\begin{tabular}{c|cccc||c|cccc}
\multicolumn{5}{c||}{$B_1$}&
\multicolumn{5}{c}{$A_1$}\cr
\hline
$g$ & ~~~~$E$~~~~ & $E_{\rm VPE}$ & ~~~~$E+E_{\rm VPE}$ & ~~$E_{\rm tot}$ &
$g$ & ~~~~$E$~~~~ & $E_{\rm VPE}$ & ~~~~$E+E_{\rm VPE}$ & ~~$E_{\rm tot}$ \cr
\hline
$2.6$ & $1.433$ & $0.714$ & $2.147$ &  $3.817$ &
$2.8$ & $2.227$ & $0.693$ & $2.920$ &  $4.456$ \\
$3.0$ & $1.370$ & $0.990$ & $2.360$ &  $4.172$ & 
$3.0$ & $2.140$ & $0.945$ & $3.091$ &  $4.812$ \\
$4.0$ & $1.039$ & $2.031$ & $3.079$ &  $5.310$ & 
$4.0$ & $2.069$ & $1.999$ & $4.068$ &  $6.147$ \\
$5.0$ & $0.715$ & $3.439$ & $4.154$ &  $6.694$ &
$5.0$ & $1.749$ & $3.687$ & $5.436$ &  $7.978$
\end{tabular}}
\caption{\label{tab:lam1}The single level fermion energies $E$ and the 
renormalized Dirac sea contributions for the case $M^2=2\lambda$ as 
a function of the Yukawa coupling $g$. Here $E_{\rm tot}=E+E_{\rm VPE}+E_{\rm cl}$.
Left panel: $B_1$ channel, right panel: $A_1$ channel.}
\end{table}

\begin{table}[h]
\centerline{
\begin{tabular}{c|cccc||c|cccc}
\multicolumn{5}{c||}{$B_1$}&
\multicolumn{5}{c}{$A_1$}\cr
\hline
$g$ & ~~~~$E$~~~~ & $E_{\rm VPE}$ & ~~~~$E+E_{\rm VPE}$ & ~~$E_{\rm tot}$ &
$g$ & ~~~~$E$~~~~ & $E_{\rm VPE}$ & ~~~~$E+E_{\rm VPE}$ & ~~$E_{\rm tot}$ \cr
\hline
$2.6$  & $1.346$  & $0.755$ & $2.101$  & $3.685$ &
$3.0$  & $2.090$  & $0.984$ & $3.074$  & $4.639$ \\
$3.0$  & $1.227$  & $1.066$ & $2.293$  & $4.058$ &
$3.5$  & $2.036$  & $1.492$ & $3.528$  & $5.327$ \\
$3.5$  & $1.026$  & $1.597$ & $2.623$  & $4.629$ &
$4.0$  & $1.875$  & $2.245$ & $4.120$  & $6.188$ \\
$4.0$  & $0.812$  & $2.237$ & $3.049$  & $5.272$ &
$4.5$  & $1.716$  & $2.990$ & $4.706$  & $6.992$ \\
\end{tabular}}
\caption{\label{tab:lam2}Same as table \ref{tab:lam1} for $M^2=1.8\lambda$.}
\end{table}
Finally, tables \ref{tab:lam1}, \ref{tab:lam2} and \ref{tab:lam3} contain our numerical results for 
the fermion contribution to the energy for configurations that include the back-reaction 
from a particular fermion level on the kink. This contribution has two components, the energy
of the explicitly occupied level and the VPE from the Dirac sea. The latter is our main novel
result. In all cases considered, the VPE adds positively and may indeed be substantial. Similar to
the classical energy, the fermion VPE differs only slightly between the $B_1$ and $A_1$ channels.
For small $g$ the VPE for the $A_1$ channel is even less than for the $B_1$ channel. This suggests 
a lesser polarization of the Dirac sea in the $A_1$ case. However, for both channels the fermion VPE 
is significantly larger than it is when occupying the $A_0$ mode. Hence the back-reaction has an
important impact on the VPE.

In table \ref{tab:lam1} we present the fermion energies in the $B_1$ and $A_1$ channels when
$M^2=2\lambda$ but for different values of the Yukawa coupling $g$. For the $B_1$ channel we find the
quasi-classical energy to obey $E+E_{\rm VPE}<g$. This suggests that the fermionic piece could be stable by
itself. However, that threshold is exceeded within the $A_1$ channel. In either case the gain in energy
from binding the single level, $g-E$, is essentially compensated by the VPE. To discuss the topological 
(in)stability we must consider the semi-classical energy $E_{\rm tot}$ which includes the Dirac sea 
contribution. We find topological stability ($E_{\rm tot}\le g+E_{\rm kink}$, with the $M^2=2\lambda$ 
values for $g+E_{\rm kink}$ listed in table \ref{tab:cl1}) merely for small~$g$ in the $B_1$ channel.
For the $A_1$ channel it does not occur at all. The data in tables \ref{tab:lam2} and \ref{tab:lam3} 
show that this picture does not change qualitatively when we vary $\lambda$ ({\it cf.} tables \ref{tab:cl2} and 
\ref{tab:cl3}, for the respective values of $g+E_{\rm kink}$). For small~$g$ the level energy increases with 
the ratio $\frac{M^2}{\lambda}$ while the VPE decreases (in the considered $\lambda$ range). As $g$ increases 
this monotonic behavior disappears.
\begin{table}
\centerline{
\begin{tabular}{c|cccc||c|cccc}
\multicolumn{5}{c||}{$B_1$}&
\multicolumn{5}{c}{$A_1$}\cr
\hline
$g$ & ~~~~$E$~~~~ & $E_{\rm VPE}$ & ~~~~$E+E_{\rm VPE}$ & ~~$E_{\rm tot}$ &
$g$ & ~~~~$E$~~~~ & $E_{\rm VPE}$ & ~~~~$E+E_{\rm VPE}$ & ~~$E_{\rm tot}$ \cr
\hline
$3.0$  & $1.472$ & $0.933$ & $2.405$  &  $4.292$ &
$3.0$  & $2.146$ & $0.942$ & $3.088$  &  $4.975$ \\
$4.0$  & $1.226$ & $1.871$ & $3.097$  &  $5.355$ &
$4.0$  & $2.050$ & $2.014$ & $4.064$  &  $6.368$ \\
$5.0$  & $0.891$ & $3.236$ & $4.127$  &  $6.736$ &
$5.0$  & $1.813$ & $3.612$ & $5.425$  &  $8.146$ \\
\end{tabular}}
\caption{\label{tab:lam3}Same as table \ref{tab:lam1} for $M^2=2.2\lambda$.}
\end{table}
The quasi-classical energy $E+E_{\rm VPE}$ is still smaller than $g$ in the $B_1$ channel for $g\le5$. 
This kinematic bound is always exceeded within the $A_1$ channel. Presumably this bound will also be exceeded 
for large enough $g$ in the $B_1$ channel because $E$ cannot become less than zero.

In Ref.~\cite{Klimashonok:2019iya} also configurations that minimize
$$
E_{\rm cl}+\epsilon_n \qquad{\rm with}\qquad \epsilon_n<0
$$
have been constructed. Even though such systems do not have a particle-hole interpretation, we have nevertheless 
computed the fermion VPE for $B_{-1}$ and $A_{-1}$ channels\footnote{The minus sign in the subscript is added
for the negative energy eigenvalue \cite{Klimashonok:2019iya}.}. These are the first available levels in the negative and 
positive parity channels. The resulting data are displayed in table \ref{tab:neg}.
\begin{table}[h]
\centerline{
\begin{tabular}{c|cc||c|cc}
\multicolumn{3}{c||}{$B_{-1}$}&
\multicolumn{3}{c}{$A_{-1}$}\cr
\hline
$g$ & $E$ & $E_{\rm VPE}$ & $g$ & $E$ & $E_{\rm VPE}$ \cr
\hline
$3.0$ & $-2.762$ & $0.500$ & $3.8$ & $-3.419$ & $0.841$   \\
$4.0$ & $-3.567$ & $0.680$ & $4.0$ & $-3.592$ & $0.892$   \\
$5.0$ & $-4.363$ & $0.864$ & $5.0$ & $-4.481$ & $1.133$   \\
\end{tabular}}
\caption{\label{tab:neg}The single level fermion energies $E$ and the 
renormalized Dirac sea contributions for the case $M^2=2\lambda$ as 
a function of the Yukawa coupling $g$. Left panel: $B_{-1}$ channel, 
right panel: $A_{-1}$ channel.}
\end{table}
For these cases the binding of the level ($E+g$) is small, but so is the VPE and there is no 
absolute gain when compared to the mass of a free fermion.

\section{Conclusion}
\label{sec:conclude}

In this study we have computed the vacuum polarization energy (VPE) emerging from fermion fluctuations
about recently observed solitons for scalar fields coupled to fermions in one time and space dimensions. 
These solutions deviate from the renowned kink by self-consistently accounting for the back-reaction from a 
single fermion bound state level. We have argued that this VPE must be considered for two reasons, at least. 
First, it equals the contribution from the Dirac sea. This contribution is needed for a physically sensible 
picture when the selected level has a negative energy eigenvalue. Second, when the model parameters are consistently 
incorporated, the VPE is of the same order in a semiclassical expansion as the energy eigenvalue of the single 
level. This was revealed by a careful analysis of the parameter dependence of the energy. Thereafter fields 
and variables could be redefined to dimensionless quantities suitable for the numerical simulation.

We have found that the fermion VPE contribution is substantial and approximately outweighs the binding of the 
single level in the parameter regime that leads to a sizable back-reaction. Nevertheless, that configuration may still 
be topologically stable for the case that the single fermion level is the first excited state and the Yukawa coupling 
is small; {\it i.e.} its total energy is less than that of a single free fermion and a kink without back-reaction. 
For moderate and large couplings this stability is lost. In other channels it never occurs. We have not found a 
configuration that is absolutely stable as the total energy has always been larger than that of a free fermion combined 
with the translationally invariant scalar configuration. Stated otherwise, the local minima of the quasi-classical
energy functional cease to exist for the semi-classical analog.

In a consistent quantum field theory calculation the Dirac sea contribution to the energy is obtained from the fermion
effective action. Its  non-local structure makes it impossible to construct the minimum of the semi-classical energy 
from a set of differential equations. Hence a variational procedure~\cite{Farhi:2000ws} that minimizes this energy 
is the most suggestive extension of the current study.

In many considerations the boson VPE may be neglected compared to the fermion one because the latter scales with 
the number of internal degrees of freedom (for example, the number of colors in soliton models for the strong
interaction). In the present model that number is unity and there is no (obvious) approximation scheme that favors 
the fermion VPE. We have seen that it strongly increases with the Yukawa coupling and for large enough coupling 
it dominates (presumably) and thus justifies the omission of the boson VPE. In any event, such a calculation would 
be even more cumbersome as a further back-reaction must then be taken into account, that of the boson fluctuations on 
the fermion wave-functions. From the historic studies in Refs. \cite{Dashen:1974ci,Ra82} we expect the boson VPE to be 
negative and to some extent cancel the fermion counterpart that we computed here. These studies also suggest a not 
so obvious scenario in which the boson VPE is negligible against the fermion counterpart: The model without fermions
also has the factor $\lambda/M^2$ for the boson VPE relative to the classical energy \cite{Graham:2009zz}. A consistent 
model in which higher order boson quantum effects are small would thus require $M^2\gg\lambda$. To nevertheless have a 
substantial back-reaction from the fermion level a large Yukawa coupling will be needed. Exploring this parameter regime 
certainly is another interesting extension of the presented study.

In Ref.~\cite{Klimashonok:2019iya} scenarios with an additional explicit mass term $\mathcal{L}_m=-m_0\overline{\Psi}\Psi$ 
were considered.  With the kink connecting the two distinct vacua at $\langle \Phi\rangle=\pm \frac{M}{\sqrt{2\lambda}}$ 
at positive and negative spatial infinity, its inclusion would lead to different mass parameters, $m_0\pm \frac{g}{2}M$,
in the fermion sector. Such systems are destabilized by quantum effects \cite{Weigel:2016zbs,Romanczukiewicz:2017hdu}.

The above discussions indicate that our estimate for the quantum corrections when kinks are
bounded to excited fermion levels leaves space for future extensions. Yet, it accounts for the important fact 
that contributions from distinct energy levels and the Dirac sea are of equal order in the semi-classical
expansion. We have therefore combined these contributions and our estimate suggests that the Dirac sea contribution
to the total energy outweighs the gain from binding single levels. This points towards an instability of such
configurations.

\acknowledgments
One of us (HW) thanks N. Graham for helpful discussions.
H.\@ W.\@ is supported in part by the National Research Foundation of
South Africa (NRF) by grant~109497.

\bibliographystyle{apsrev}

\end{document}